\documentclass[aps,prl,amsmath,amssymb,preprint]{revtex4-2}
\usepackage{graphicx}
\graphicspath{{figures/}}
\usepackage{color}
\usepackage{float}
\usepackage{soul}
\usepackage{dcolumn}
\usepackage{bm}
\usepackage{xcolor} 
\usepackage[unicode,colorlinks=true]{hyperref}
\usepackage{upgreek}
\usepackage{siunitx}
\begin{document}

\section*{Title}
Ultrafast Opto-Electronic and Thermal Tuning of Third-Harmonic Generation in a Graphene Field Effect Transistor


\section*{Author list}
Omid Ghaebi$^{1}$, Sebastian Klimmer$^{1,2}$, Nele Tornow$^{1}$, Niels Buijssen$^{1}$, Takashi Taniguchi$^{3}$, Kenji Watanabe$^{4}$, Andrea Tomadin$^{5}$, Habib Rostami$^{6}$, Giancarlo Soavi$^{1,7}$

\section*{Affiliations}
$^1$Institute of Solid State Physics, Friedrich Schiller University Jena, Jena, Germany

$^2$ARC Centre of Excellence for Transformative Meta-Optical Systems, Department of Electronic Materials Engineering, Research School of Physics, The Australian National University, Canberra, Australia

$^3$Research Center for Materials Nanoarchitectonics, National Institute for Materials Science,  1-1 Namiki, Tsukuba 305-0044, Japan

$^4$Research Center for Electronic and Optical Materials, National Institute for Materials Science, 1-1 Namiki, Tsukuba 305-0044, Japan

$^5$Dipartimento di Fisica, Università di Pisa, Largo Bruno Pontecorvo 3, 
56127 Pisa, Italy

$^6$Department of Physics, University of Bath, Claverton Down, Bath BA2 7AY, United Kingdom

$^7$Abbe Center of Photonics, Friedrich Schiller University Jena, Jena, Germany

\section*{Abstract}
Graphene is a unique platform for tunable opto-electronic applications thanks to its linear band dispersion, which allows electrical control of resonant light-matter interactions. Tuning the nonlinear optical response of graphene is possible both electrically and in an all-optical fashion, but each approach involves a trade-off between speed and modulation depth. Here, we combine lattice temperature, electron doping, and all-optical tuning of third-harmonic generation in a hBN-encapsulated graphene opto-electronic device and demonstrate up to 85\% modulation depth along with gate-tunable ultrafast dynamics. These results arise from the dynamic changes in the transient electronic temperature combined with Pauli blocking induced by the out-of-equilibrium chemical potential. Our work provides a detailed description of the transient nonlinear optical and electronic response of graphene, which is crucial for the design of nanoscale and ultrafast optical modulators, detectors and frequency converters.

\maketitle
\section{Introduction}
Two-dimensional (2D) materials are ideal candidates for nonlinear optical applications at the nanoscale~\cite{dogadov2022parametric}, as they enable ultra-broadband optical parametric amplification~\cite{trovatello2021optical}, spontaneous parametric down-conversion~\cite{guo2023ultrathin}, electrical and all-optical tuning of the second harmonic  (SH)~\cite{klimmer2021all,cheng2020ultrafast,seyler2015electrical,taghinejad2020photocarrier} and third harmonic (TH) generation~\cite{cheng2020ultrafast,soavi2018broadband,soavi2019hot}, giant efficiencies of THz high harmonic generation~\cite{hafez2018extremely}, and applications in integrated nonlinear opto-electronic devices such as gas sensors~\cite{an2020electrically}, logic gates~\cite{li2022nonlinear,zhang2022chirality} and valleytronics ~\cite{herrmann2023nonlinear,herrmann2023valley}. 

Within the family of 2D materials, graphene arguably shows the most intriguing nonlinear response. Being centrosymmetric, the first nonlinear term in its polarization is the third-order susceptibility $\chi^{(3)}$. While few experimental studies have observed SHG due to breaking of symmetry at an interface~\cite{dean2010graphene,dean2009second}, in-plane electric fields and currents~\cite{bykov2012second,an2013enhanced} or from the electric quadrupole response ~\cite{zhang2019doping}, the vast majority of nonlinear optical experiments on graphene have focused on $\chi^{(3)}$ processes such as FWM ~\cite{jiang2018gate}, THG~\cite{soavi2018broadband,alonso2021giant,kumar2013third,hong2013optical,soavi2019hot} and saturable absorption~\cite{kumar2009femtosecond,sun2010graphene,malouf2019two}. In particular, THG and FWM have recently gained increasing attention following the demonstration of their electrical~\cite{jiang2018gate,soavi2018broadband,soavi2019hot} and all-optical ~\cite{cheng2020ultrafast} modulation, which provide a route towards ultrafast nanoscale frequency converters and a powerful method to probe ultrafast hot electron dynamics. The electrical tunability of THG in graphene has been widely explored~\cite{soavi2018broadband,soavi2019hot,alonso2021giant,jiang2018gate}, whereas the interplay of lattice and electron temperatures in high-quality hBN-encapsulated (hexagonal boron nitride) graphene samples is scarcely studied. In addition, in the case of all-optical modulation, only one experiment in the visible and UV frequencies (THG centered at $\sim$\;\SI{450}{\nano\metre} and thus outside the Dirac cone) has been reported to date~\cite{cheng2020ultrafast}. There, Cheng \textit{et al.} have shown all-optical TH modulation depth up to 90$\%$ for pump fluences of \SI{40}{\milli\joule\per\square\centi\metre} (excitation wavelength of \SI{400}{\nano\metre}) and \SI{25}{\milli\joule\per\square\centi\metre} (excitation wavelength of \SI{800}{\nano\metre}) with a relaxation time-constant of $\sim$\;\SI{2.5}{\pico\second}. Further, all-optical THG modulation was attributed solely to Pauli blocking, while the role of the electronic temperature ($T_{\rm e}$) and its impact on the $\chi^{(3)}$ tensor was neglected. 

In this work, we provide a detailed experimental and theoretical study of ultrafast thermal and opto-electronic modulation of THG in a high-quality and gate-tunable hBN/graphene/hBN field effect transistor (FET). Our scheme for opto-electronic THG modulation can be briefly summarized as follows. We irradiate graphene with two pulses: a fundamental beam (FB) and a control beam (CB). The FB is responsible for inducing the parametric THG process ($\omega_{\rm FB}$ $\rightarrow$ $3\rm\omega_{\rm FB}$) while the CB controls the THG efficiency \textit{via} tuning of $T_{\rm e}$ and Pauli blocking. We point out from the very start that the FB affects $T_{\rm e}$ and Pauli blocking as well, due to its large fluence (comparable to the CB), necessary to generate a sizable THG. Furthermore, electrical doping by means of external gates enables the system to modulate the competition between $T_{\rm e}$ and Pauli blocking mechanisms and to tune the THG ultrafast recombination dynamics. Thus, by combining electrical and all-optical control of $T_{\rm e}$ and $E_{\rm F}$ we achieve active modulation of THG in graphene with the following main results. First, experiments on encapsulated samples allow to show that the electrical modulation of THG in graphene is symmetric for electrons and holes within the Dirac cone. This is the nonlinear optical analog of the electronic ambipolar behaviour of FETs which was absent in previous studies~\cite{soavi2018broadband,jiang2018gate}. Further, we observe up to 300\% modulation in the THG intensity by tuning the lattice temperature ($T_{\rm L}$) from \qtyrange[range-units=single]{295}{33}{\kelvin}. Second, we show that electrical doping can be used to actively control the recombination dynamics of the THG signal arising from phase-space quenching of the scattering between hot electrons and optical phonons~\cite{pogna2021hot}. Third, we shed light on the physical origin of the ultrafast THG modulation and the interplay of hot electrons and Pauli blocking. Finally, with our nonlinear opto-electronic device, we achieve a TH modulation depth of $ \approx$\;85\% at $E_{\rm F}=$\;\SI{300}{\milli\electronvolt} and peak fluence of \SI{200}{\micro\joule\per\square\centi\metre}. Comparing this result to the only all-optical THG modulation of graphene reported to date~\cite{cheng2020ultrafast}, we achieve a similar modulation depth at an excitation fluence that is more than two orders of magnitude lower. This is possible thanks to mid-IR excitation and active control of $E_{\rm F}$ and $T_{\rm L}$ and thus it further clarifies that a deeper understanding of the ultrafast and nonlinear opto-electronic response of graphene is paramount for the design and optimization of nanoscale ultrafast devices, such as optical modulators, detectors, and frequency converters.

\section{Ambipolar gate-tunable THG}

\begin{figure*}
\centering
\includegraphics[width=\linewidth]{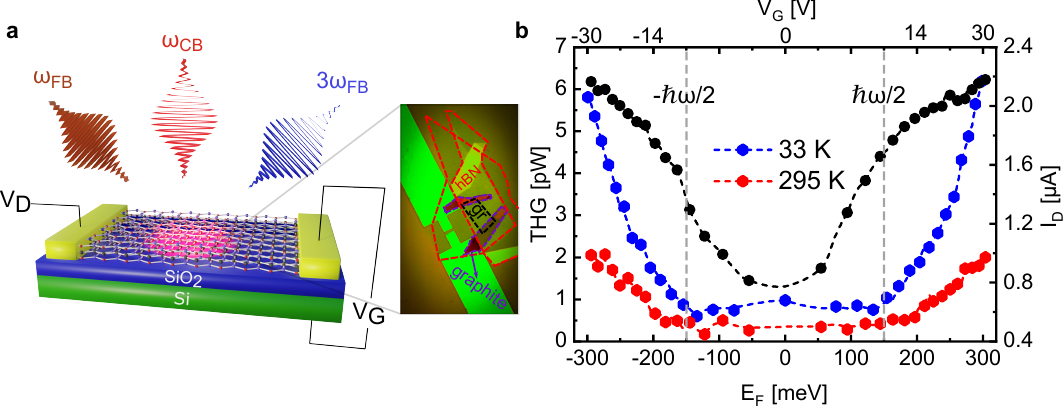}
\caption{\label{fig:ambipolar}
\textbf{Opto-electronic modulation of THG in a graphene FET.} \textbf{a)} Sketch and microscope optical image of the device. Monolayer graphene is encapsulated between two hBN flakes. $V_{\rm G}$, $V_{\rm D}$, $\omega_{\rm FB}$ and $\omega_{\rm CB}$ represent the gate-source voltage, source-drain voltage, fundamental beam, and control beam, respectively. \textbf{b)} THG as a function of $E_{\rm F}$ (bottom x-axis) and $V_G$ (top x-axis) at lattice temperatures of $T_{\rm L}=$\;\SI{295}{\kelvin} (red curve) and $T_{\rm L}=$\;\SI{30}{\kelvin} (blue curve). The black curve is the drain current ($I_{D}$) as a function of $E_{\rm F}$ and $V_{\rm G}$ at the drain voltage of $V_{\rm D}= 1$ mV.
}
\end{figure*}

Opto-electronic (\textit{i.e.}, ~optical and electrical) modulation of THG is performed on a back-gated FET based on a single layer graphene encapsulated in two $\sim$\;\SI{10}{\nano\metre} thick hBN layers (Fig.\;\ref{fig:ambipolar}a). The device was prepared by mechanical exfoliation and dry transfer, following the approach described in Ref.~\cite{purdie2018cleaning} (see Supplementary Information S1 and S2 for details on sample fabrication and characterization). For the THG measurements, we used two synchronized laser pulses at photon energies of \SI{0.32}{\electronvolt} (\SI{3900}{\nano\metre}) and \SI{1.2}{\electronvolt} (\SI{1030}{\nano\metre}) for the FB and CB, respectively (see Supplementary Information S3). 

First, we measured gate-tunable THG with a ``static'' procedure (\textit{i.e.}, without CB). We irradiate our device with the FB (\SI{130}{\micro\joule\per\square\centi\metre}) and collect the TH power for different values of the applied $V_{\rm G}$ in the range \qtyrange[range-units=single]{-30}{30}{\volt}, corresponding to values of the $E_{\rm F}$ in the range \qtyrange[range-units=single]{-300}{300}{\milli\electronvolt} (see Supplementary Information S2 for the calculation of $E_{\rm F}$) and for different $T_{\rm L}$. The experimental data (Fig.\;\ref{fig:ambipolar}b) show a modulation factor of $\sim$\;4  when $T_{\rm L}=$\;\SI{295}{\kelvin} and the $E_{\rm F}$ is tuned from $\sim$\;\SI{50}{\milli\electronvolt} to \SI{300}{\milli\electronvolt}.
This gate-tunable TH modulation is due to the crossing of multi-photon resonances in the Dirac cone, as largely discussed in Refs.~\cite{soavi2018broadband,soavi2019hot}.
Once the $T_{\rm L}$ is decreased to \SI{33}{\kelvin}, the modulation factor in the same $E_{\rm F}$ range increases to $\sim$\;9. 
Comparing the two curves at different temperatures, we observe an enhancement of the TH power while reducing $T_L$ of $\sim$\;1.5 and $\sim$\;3 at $E_{\rm F}=$\;\SI{50}{\milli\electronvolt} and $E_{\rm F}= $\;\SI{300}{\milli\electronvolt}, respectively. The origin of this remarkable enhancement of THG with lattice temperature is manifold.
Our theoretical analysis reproduces this effect, on a smaller magnitude, solely based on the different electron distribution achieved when samples with different $T_{\rm L}$ are irradiated by the same FB. 
This is an indirect effect of $T_L$ on THG, due to the different dynamics experienced by electrons on a statistical level.
However, we assume that a contribution to the observed TH enhancement arises also from a direct effect of temperature at the level of single-particle, coherent evolution during the FB pulse duration. Such an effect can be attributed to the temperature-dependent electron scattering rates (or electron spectral broadening) with impurities, defects and phonons (see Supplementary Information S5).
Although our numerical calculations support this argument, a solid determination of the scattering rates at different temperatures would require a much larger amount of data sets which is outside the scope of this work. 

The absence of sharp peaks in the data reported in Fig.\;\ref{fig:ambipolar}b is a clear indication of the high $T_{\rm e}$ reached during the experiments ~\cite{soavi2018broadband,soavi2019hot}, as we discuss in detail in the Supplementary Information S4.
Since $T_{\rm e}$ is a function of $E_{\rm F}$ and varies dramatically over the pulse duration, we cannot assign a single value of $T_{\rm e}$ to the points in Fig.\;\ref{fig:ambipolar}b. However, if we consider, \textit{e.g.} $T_{\rm L} =$\;\SI{33}{\kelvin} and $E_{\rm F}=$\;\SI{50}{\milli\electronvolt}, our calculations show that a $T_{\rm e}$ $>$ \SI{1400}{\kelvin} is achieved by the electron distribution for over \SI{200}{\femto\second}, at the FB peak fluence of \SI{130}{\micro\joule\per\square\centi\metre} (see also Supplementary Information S4). We point out that we observe gate-tunable THG for both positive and negative values of the $E_{\rm F}$, indicating that the THG enhancement at multi-photon resonances can be achieved for both n- and p-doping \textit{i.e.} in the conduction and valence band of the Dirac cone, qualitatively preserving the electron-hole symmetry of the phenomenon to a remarkable degree.

Finally, the results reported in Fig.\;\ref{fig:ambipolar}b allow us to estimate the $\chi^{(3)}$ of graphene at different values of $E_{\rm F}$, at the FB photon energy of \SI{0.32}{\electronvolt} by using the two following equations ~\cite{alonso2021giant}:

\begin{equation}\label{eq:power_chi3}
  P({\omega }_{i,o})=\frac{1}{8}{\left(\frac{\pi }{\mathrm{ln}\,2}\right)}^{3/2}f\tau {W}^{2}{n}_{{\omega }_{i,o}}{\epsilon }_{0}c\frac{| E({\omega }_{i,o}){| }^{2}}{2}
\end{equation}

\begin{equation}\label{eq:E_chi3}
   E({\omega }_{{{o}}})=\frac{1}{4}\frac{{{i}}{\omega }_{{{i}}}}{2\pi c}{\chi }_{\exp }^{(3)}{d}_{{\rm{gr}}}{E}^{3}({\omega }_{{{i}}})
\end{equation}

where $P({\omega }_{i,o})$, $E({\omega }_{i,o})$ are the input/generated THG power and electric field and f,$\tau$, ${n}_{{\omega }_{i,o}}$ are the repetition rate, pulse duration, and refractive index, respectively. The input/THG electric fields can be extracted from equation~(\ref{eq:power_chi3}) and then the $\chi^{(3)}$ value can be calculated using equation~(\ref{eq:E_chi3}). ${d}_{\rm{gr}}=$\;\SI{0.3}{\nano\metre} is the thickness of monolayer graphene.
Considering the losses of the setup (see Supplementary Information S3) and $T_{\rm L} =$\;\SI{33}{\kelvin}  we obtain $\chi^{(3)} \sim$\;\SI[print-unity-mantissa=false]{2e-15}{\metre\squared\per\volt\squared} for $E_{\rm F}\sim$\;\SI{300}{\milli\electronvolt} and $\sim$\;\SI[print-unity-mantissa=false]{8e-16}{\metre\squared\per\volt\squared} for $E_{\rm F}\sim$\;\SI{0}{\milli\electronvolt}, in agreement with Ref.~\cite{alonso2021giant} where a $\chi^{(3)}\sim$\;\SI[print-unity-mantissa=false]{6e-16}{\metre\squared\per\volt\squared} was reported for pristine graphene at a fundamental photon energy of \SI{0.225}{\electronvolt} and $E_{\rm F}=$\;\SI{390}{\milli\electronvolt}.

\begin{figure*}
\centering
\includegraphics[width=\linewidth]{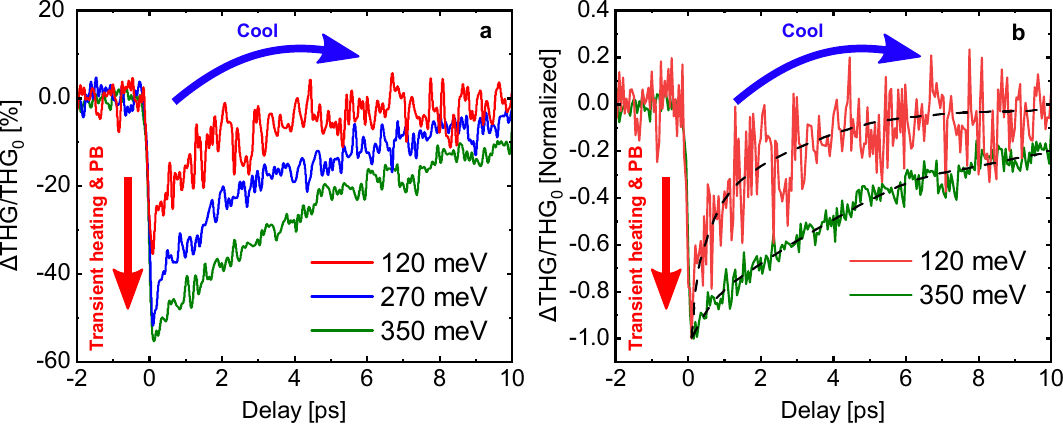}
\caption{\label{fig:relaxation}
\textbf{All-optical modulation of THG and gate tunable dynamics.}
\textbf{a)} Ratio \( \frac{\Delta THG}{THG_{0}} \), (defined in equation~(\ref{eq:delta_thg})), for different values of $E_{\rm F}$. ${THG_{0}}$ has been measured at \SI{-2}{\pico\second}. The interplay of transient heating and PB (Pauli blocking) on electrons will occur when CB and FB pulses are spatially and temporally synchronized and subsequent cooling occurs \textit{via} electron-electron and electron-phonon scattering. \textbf{b)} Normalized \( \frac{\Delta THG}{THG_{0}} \) for $E_{\rm F}=$\;\SI{120}{\milli\electronvolt} and $E_{\rm F}=$\;\SI{350}{\milli\electronvolt}.}
\end{figure*} 

\section{Ultrafast opto-electronic TH modulation}
Next, we shift our attention to time-resolved and all-optical TH modulation. We initially fix the CB and FB fluence at \qtylist[list-units=single]{170;110}{\micro\joule\per\square\centi\metre}, respectively, and scan their relative delay for different values of $E_{\rm F}$ in the range \qtyrange[range-units=single]{0}{390}{\milli\electronvolt}. We remark that this range of $E_F$ overlaps the region defined by the lower threshold $E_{\rm F}> \hbar\omega / 2$, where absorption of the FB, at zero temperature, is forbidden by Pauli blocking. However, we do not see an abrupt drop-off of the measured signal when the Fermi energy exceeds such threshold.
The reason is that the finite temperature in our samples ensures that a residual absorption is always present.
Even a small initial absorption produces a rapid temperature increase, which broadens the electron distribution in the energy space and relaxes the condition for Pauli blocking. To mitigate the effect of diminished absorption, in the following, we discuss the behavior of the measured signal divided by the signal before the pump is applied, thus ``normalizing-out'' the most trivial part of the Pauli blocking. We point out, however, that other non-trivial thresholds appear in the THG as the Fermi energy crosses multiples of the FB frequency~\cite{soavi2018broadband,soavi2019hot}. Fig.\;\ref{fig:relaxation} shows the experimental results for the ratio $\Delta THG / THG_{0}$, where
\begin{equation}\label{eq:delta_thg}
\Delta THG (\tau) = THG (\tau) - THG_{0}~,
\end{equation}
$THG (\tau)$ is the measured signal as a function of delay $\tau$, and $THG_{0}$ is the reference THG measured in the absence of the CB, that we measure at a negative delay $\tau =$\;\SI{-2}{\pico\second}.
As expected, the signal features a sharp peak when the FB overlaps with the CB, \textit{i.e.} when both beams excite the electron system, followed by a ``relaxation'' stage converging to a zero signal, which represents the recovery of the system from the excitation due to the CB. At large delays, the effect of the CB vanishes and the THG recovers to its reference value $THG_{0}$.

The process of electron relaxation in graphene after excitation from an ultrashort pulse has been discussed at length in the literature ~\cite{kadi2015impact, tielrooij2013photoexcitation, brida2013ultrafast, tomadin2013nonequilibrium, massicotte2021hot, Pogna_acsnano_2022} and, in summary, involves: (i) an initial stage dominated by electron-electron interactions where the photoexcited electron system achieves thermalization at a temperature much higher than the initial (lattice) temperature, possibly with inter-band processes associated to Auger recombination and carrier multiplication; (ii) a first cooling stage dominated by the emission of optical phonons where both the electron temperature and the photoexcited density decreases; (iii) a second, slower cooling stage, where the hot optical phonons thermalize with the acoustic phonons of the lattice, possibly with the intervention of ``supercollision'' processes, and the unperturbed initial state is finally recovered.
We remark again that the FB, due to its fluence, strongly perturbs the electron system, such that, even several \SI{}{\pico\second} after the CB, the THG signal cannot be considered as the response of an electron system at equilibrium with the lattice. 

From the data in Fig.\;\ref{fig:relaxation}, we also notice that the rate of relaxation diminishes as the Fermi energy is increased.
We recognize this effect as the quenching of optical phonon emission in the first cooling stage, due to the reduction of the available phase-space for electronic transitions, which was recently discussed in Ref.~\cite{pogna2021hot}.
In other words, due to Pauli blocking, photoexcited electrons at energy $E$ can only emit a phonon of energy $\hbar \omega_{\rm ph}$, if states are available at energy $E - \hbar \omega_{\rm ph}$.
As the Fermi energy is increased, and approaches the photoexcitation energy, this condition is harder and harder to satisfy, even at large temperature where the electron distribution is broadened.
It is interesting that this phase-space effect does not only affect the differential transmission of the electron system, as demonstrated in Ref.~\cite{pogna2021hot}, but emerges in the measurement of the THG as well. 
This observation highlights how consequential it is to be able to tune the electron density by electrical doping in a graphene-based optoelectronic device, thus exerting a certain degree of control on both its linear and non-linear optical response. 

\begin{figure*}
\centering
\includegraphics[width=\linewidth]{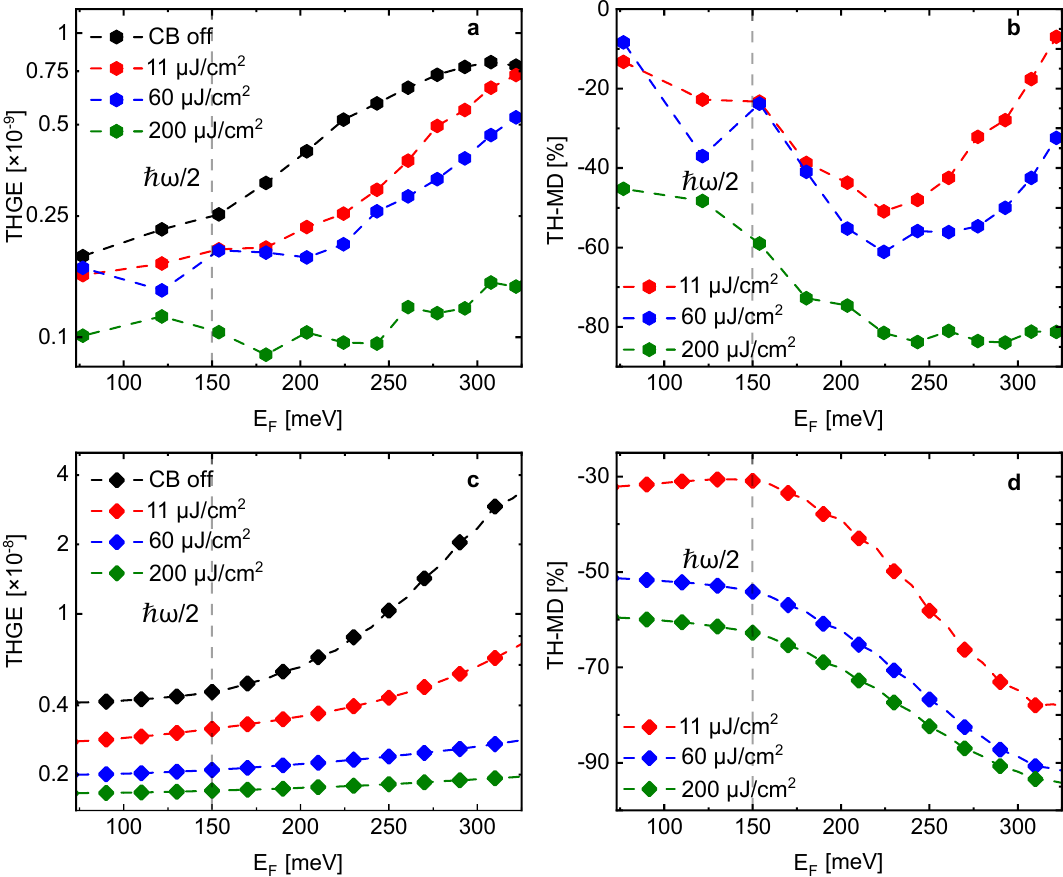}
\caption{\label{fig:modulation}
\textbf{Influence of $E_{\rm F}$ and CB peak fluence on THGE and {TH-MD}} \textbf{a,b)} Experimental THGE and {TH-MD} (defined in equation~(\ref{eq:exp_obs})) as a function of $E_{\rm F}$, for different values of the CB peak fluence reported in the legend. \textbf{c,d)} Theoretical THGE and {TH-MD} calculated using the experimental values of the incident peak fluences of CB and FB.} 
\end{figure*} 

Finally, we explore the dependence of the THG on the state of the electron system before the FB, by changing the fluence of the CB. In Fig.\;\ref{fig:modulation}a we plot the THG efficiency (THGE) and in Fig.\;\ref{fig:modulation}b the third harmonic modulation depth ({TH-MD}), defined as
\begin{equation}\label{eq:exp_obs}
{\rm THGE} = \frac{P_{TH}}{P_{FB}}~,  \quad
{\rm {TH-MD}} = \frac{\Delta P_{TH}}{P_{TH_0}}~,
\end{equation}
respectively, where $\Delta P_{TH}$ is the difference in the TH power ($P_{TH}$) with and without ($P_{TH_0}$) the CB, and $P_{FB}$ is the power of the fundamental beam.
The data are shown as a function of $E_{\rm F}$ and for different values of the incident CB fluence. In all the experimental graphs, the data are extracted at zero time delay between the FB and CB. 

When the CB is off (black symbols in Fig.\;\ref{fig:modulation}a), we obtain a similar result reported in Fig.\;\ref{fig:ambipolar}b, namely an increase of the THGE when $\hbar\omega<2E_{\rm F}$. The same trend can be observed when we switch-on the CB, but in this case, the modulation factor with respect to $E_{\rm F}$ is reduced. When the CB fluence reaches \SI{200}{\micro\joule\per\square\centi\metre} (green symbols) the modulation factor is close to zero and the THGE is almost constant over the measured range of $E_{\rm F}$. 

The {TH-MD} is in the range $\sim$\;\qtyrange[range-units=single]{7}{85}{\%} for CB peak fluences of \qtyrange[range-units=single]{11}{200}{\micro\joule\per\square\centi\metre}. Interestingly, we obtain a maximum {TH-MD} of 85\% for $E_{\rm F}=$\;\SI{300}{\milli\electronvolt} and peak fluence of \SI{200}{\micro\joule\per\square\centi\metre}. This exceeds by far the results of Ref.~\cite{cheng2020ultrafast}, where a similar {TH-MD} of 90\% was obtained for a CB peak fluence of \SI{25}{\milli\joule\per\square\centi\metre}.
Two features of the data deserve to be highlighted: (i) tuning $E_{\rm F}$ plays a huge role in the {TH-MD}; (ii) for all values of $E_{\rm F}$ we observe a negative {TH-MD}.

Fig.\;\ref{fig:modulation}c and d show our theoretical calculations for the THGE and {TH-MD}, respectively, obtained by means of the model discussed in the following section.
The overall agreement between theory and experiment is satisfactory, albeit with two shortcomings. The first is an overall factor in the magnitude of the signal, which can easily be traced back to an incomplete determination of some fitting parameters, such as the attenuation of the signal in the detection apparatus, or the electron scattering rates in the theoretical expression of the THG (see Supplementary Infomation S5).
The second is the missing ramp-up of the {TH-MD} at $E_{\rm F} \gtrsim$\;\SI{250}{\milli\electronvolt}.
We find this discrepancy similar to what was reported in Ref.~\cite{Pogna_acsnano_2022} in the context of the quenching of the optical phonon-emission by Pauli blocking and attribute it to the theoretical model missing a Fermi-energy-dependent effect which enhances electron recombination.
In any case, these two shortcomings do not hinder our understanding of the main feature which we are concerned with in the present work, namely the all-optical switching of the TH signal. The theoretical results fully support our picture that the variations of the measured signal are due to the effect of the CB on the electron distribution before the sample is irradiated by FB.

\section{Theory of ultrafast opto-electronic THG modulation}
\label{sec:theory}

\subsection{THG efficiency for photoexcited electrons}

In order to rationalize our experimental results, we need to extend the theoretical treatment of the THG~\cite{soavi2018broadband,soavi2019hot} to take into account the specific role that the CB plays in the dynamics of the electron system. Indeed, the key issue of the CB-FB protocol used in our experimental procedure is that the increase of $T_e$, due to the heat delivered by the CB, is inextricably linked to the production of a photoexcited electron density $\delta n_{\rm e}$, \textit{i.e.},~an excess electron (hole) density in the conduction (valence) band. We emphasize that such an excess carrier density is larger than the density that appears in an equilibrium system when the temperature is increased, purely due to the broadening of the Fermi-Dirac distribution across the Dirac point. Mathematically, $\delta n_{\rm e}$ results in the splitting of the chemical potential $\mu$ into two different chemical potentials $\mu_{\rm C}$, $\mu_{\rm V}$ for the electrons in conduction and valence bands, respectively, also known as ``quasi-Fermi energies''.
We emphasize that the proper $E_{\rm F}$, an equilibrium quantity that corresponds to the value of the chemical potential at vanishing temperature, is in a one-to-one correspondence to the electron density due to doping, and does not change due to process of inter-band photoexcitation.

Following Refs.~\cite{soavi2018broadband,soavi2019hot}, it is convenient to factor equation~(\ref{eq:exp_obs}) for THGE as
\begin{equation}\label{eq:etaTHG}
THGE= \frac{n_{\rm b}}{n_{\rm t}^{3}(n_{\rm t}+n_{\rm b})^2} \left(\frac{I _{\rm FB}}{W_0}\right)^2 |S(\omega_{\rm FB}+i\Gamma_e,\mu_{\rm C},\mu_{\rm V},T_e)|^2~,
\end{equation}
where $n_{\rm t}$, $n_{\rm b}$ are the refractive indices of the top and bottom substrates, respectively, and the quantity $W_0=10^{12} {\rm W/m^2}$ is introduced to render the expression dimensionless.
Finally, the factor $S$ is the TH conductivity, which depends on the frequency $\omega_{\rm FB}$ of the FB pulse and on the thermodynamic variables of the photoexcited electron system, \textit{i.e.}, $T_e$ and the two chemical potentials $\mu_{\rm C}$ and $\mu_{\rm V}$. The expression for the TH conductivity at zero temperature ($T_{\rm e}=0$), in the absence of photoexcited density ($\delta n_{\rm e} = 0$, i.e.~$\mu_{\rm C}=\mu_{\rm C}=\varepsilon_{\rm F}$), was given in Ref.~\cite{Rostami_prb_2016} in a fully analytical form, and reads
\begin{equation}\label{eq:S_equil}
S(\hbar\omega_{\rm FB}+i\Gamma_e,E_{\rm F}) = K(E_{\rm F})  \frac{17 G(X/2) -64 G(X) + 45 G(3X/2)}{X^4}~,
\end{equation}
in terms of the dimensionless function $G(X)=\ln\left[(1+X)/(1-X)\right]$.
The parameter $K$ is a dimensionless constant given by 
\begin{equation}\label{eq:const_K}
K(E_{\rm F})= \frac{W_0}{2\epsilon^2_0 c^2} \frac{e^4\hbar v^2_{\rm F}}{192\pi E_{\rm F}^{4}}~.
\end{equation}
Finally, the dimensionless quantity $X= (\hbar\omega_{\rm FB}+i\Gamma_{\rm e})/|E_{\rm F}|$ in equation~(\ref{eq:S_equil}) is the energy of the FB photons, rescaled by $E_{\rm F}$, and includes an imaginary contribution due to the effective electron scattering rate $\Gamma_{\rm e}$. The expression of $\Gamma_{\rm e}$ depends on the precise scattering channel responsible for the finite electron mobility, such as charged impurities, phonon, defects \textit{etc.}, and it might depend on the electron doping as well as the electron and lattice temperatures (see Supplementary Information S5).

\begin{figure}
\includegraphics[width=\linewidth]{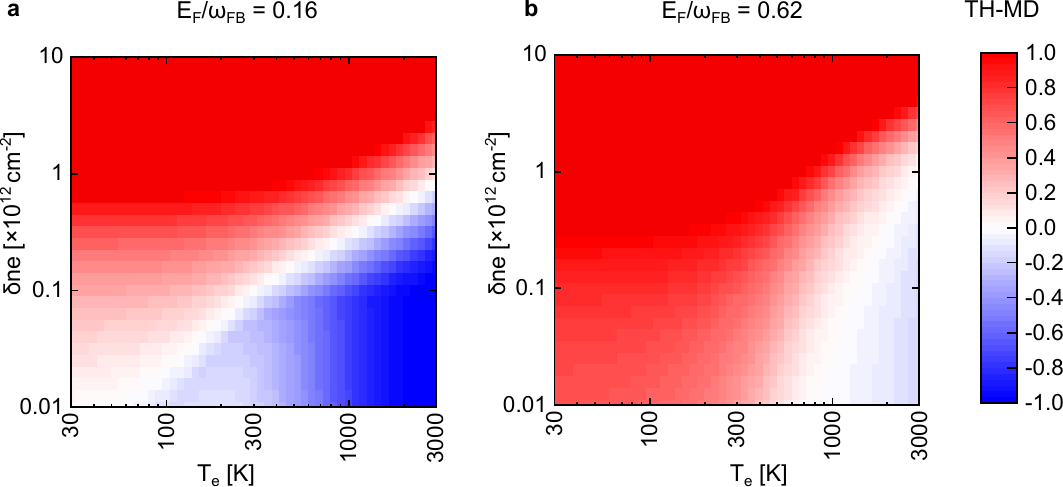}
\caption{\label{fig:efficiency}
{\bf {TH-MD} of the photoexcited electron system.}
The calculated value of the {TH-MD} as a function of $T_e$ and $\delta n_{\rm e}$ at fixed $E_{\rm F}$ equal to \textbf{(a)} $0.16$ and \textbf{(b)} $0.62$ in units of the FB frequency $\omega_{\rm FB}$. At a fixed $T_e$, increasing the value of $\delta n_e$ will enhance the THG signal, while at a fixed $\delta n_e$, increasing $T_e$ will reduce THG intensity. Both $T_e$ and $\delta n_e$ can be controlled by the incident laser power absorption.}
\end{figure}

To obtain the expression of the TH conductivity of the photoexcited electron gas, we now apply a well-known algebraic trick due to Maldague~\cite{giuliani2005quantum}, as detailed in Ref.~\cite{Tomadin_prb_2013} for the linear polarization function (\textit{i.e.},~the Lindhard function). This approach allows us to calculate the desired quantity numerically, by means of an energy-integral over the analytical expression given in equation~(\ref{eq:S_equil}):
 \begin{align}\label{eq:S_photoexc}
S(\hbar\omega_{\rm FB}+i\Gamma_e, \mu_{\rm C},\mu_{\rm V},T_e) & = \frac{1}{4k_{\rm B}T_e} 
\int^\infty_0 dE \left\{ 
\frac{S\left(\hbar\omega_{\rm FB}+i\Gamma_e,E\right)}{\cosh^2\left( \frac{E-\mu_{\rm C}}{2k_{\rm B}T_e} \right)} 
+
\frac{S\left(\hbar\omega_{\rm FB}+i\Gamma_e,E\right)}{\cosh^2\left( \frac{E+\mu_{\rm V}}{2k_{\rm B}T_e} \right)} 
\right\}  
\nonumber\\&
- \left\{\frac{1}{e^{-\mu_{\rm C}/k_{\rm B}T_e}+1} 
- \frac{1}{e^{-\mu_{\rm V}/k_{\rm B}T_e}+1} \right\} S\left(\hbar\omega_{\rm FB}+i\Gamma_e,E_{\rm F}\to0\right)~.
\end{align}
The standard mathematical expression that relates the $\mu_{\rm V}$, $\mu_{\rm C}$ the $\delta n_{\rm e}$, and $E_{\rm F}$ can be found \textit{e.g.}~in Ref.~\cite{soavi2018broadband}.

To better illustrate the dependence of the THG on a {\it variation} of electron temperature and photoexcited electron density, in Fig.\;\ref{fig:efficiency} we show the profile of the {TH-MD} (defined in equation~(\ref{eq:exp_obs})), with respect to a reference state with $T_{\rm e} = T_{\rm L}$ and vanishing $\delta n_{\rm e}$.
As expected from the equilibrium results~\cite{Rostami_prb_2016}, increasing the $T_e$ generally lowers the value of the $P_{TH}$ (\textit{i.e.}, negative {TH-MD}). Increasing the $\delta n_{\rm e}$, on the contrary, increases the $P_{TH}$, as can also be expected from the doping-dependence known from the equilibrium results~\cite{Rostami_prb_2016}. In other words, $\delta n_{\rm e}$ can be seen as a quasi-equilibrium electron- and hole-doping in conduction and valence band, respectively.
It follows that the CB can affect the THG in two {\it competing} ways because it produces a $T_e$ increase that is necessarily coupled to the production of $\delta n_{\rm e}$. It is then necessary to know the precise relation between $T_{\rm e}(t)$ and $\delta n_{\rm e}(t)$ {\it in time} to predict the THG following a given CB. To this end, we resort to the solution of a model dynamics, based on a simple rate-equation approach, which we outline in the following section.

Before we discuss our dynamical model, we remark that the procedure that leads to equation~(\ref{eq:S_photoexc}) cannot be applied to arbitrary non-equilibrium states of the electron system, but assumes that the carriers in the two bands are thermalized to the same $T_{\rm e}$, although it allows for two different $\mu_{\rm V}$ and $\mu_{\rm C}$. Mathematically, this means that the electron (hole) distribution in the conduction (valence) band is given by a Fermi-Dirac function of the form
\begin{align}\label{eq:quasiequilibrium}
f_{\rm e,h}(E,\mu_{\rm C,V}(t),T_e(t)) = \frac{1}{e^{(E \pm \mu_{\rm C,V}(t))/k_{\rm B} T_e(t)}+1}~,
\end{align}
where the carrier energy $E$ is measured from the Dirac point.The quasi-equilibrium assumption of equation~(\ref{eq:quasiequilibrium}) then holds if the system's dynamics is coarse-grained on a time-scale longer than the electron thermalization time-scale, which has been shown to be shorter than $\sim$\;\SI{20}{\femto\second} in graphene~\cite{brida2013ultrafast}. The dynamical model that we adopt here is fully consistent with this limitation.

\subsection{Model dynamics of photoexcited electrons}

To model the dynamics of photoexcited electrons, we adopt a rate-equation approach that describes: (i) electron heating due to the laser beams; (ii) energy exchange between electrons and optical phonons, due to emission and absorption processes; (iii) optical phonon relaxation to the lattice equilibrium temperature (see \textit{e.g.}~Ref.~\cite{Pogna_acsnano_2022} and references therein).
The variables of interest are the  $T_{\rm e}(t)$, $\delta n_{\rm e}(t)$, and the occupation of the optical phonon modes around the center of the Brillouin zone ($\Gamma$ point) and the valleys ($K$ points), with frequencies $\omega_\Gamma$ and $\omega_K$, respectively.

The time-derivative of the $T_e$ is given by the net absorbed power divided by the heat capacity
\begin{equation}\label{eq:eom_temp}
\frac{dT_{\rm e}(t)}{dt} = \frac{{\cal P}(t) - R_{\Gamma}(t) \hbar \omega_{\Gamma} - R_{K}(t) \hbar \omega_{K}}{c_{\rm e}(t) + c_{\rm h}(t)}~,
\end{equation}
where ${\cal P}(t)={\cal P}_{\rm FB}(t)+{\cal P}_{\rm CB}(t)$ is the average power absorbed per unit area, $c_{\rm e,h}(t)$ are the electron and hole heat capacity per unit area, and $R_{\Gamma,K}(t)$ are the net phonon emission and absorption rates.
The expressions for the electron absorbance (which relates the absorbed to the incident power in the linear regime) and the heat capacity can be found \textit{e.g.}~in Ref.~\cite{soavi2018broadband}. Here, we calculate the heat capacity as the sum of the electron and hole contribution, taken into account independently, because inter-band recombination processes are much slower than thermalization, and thus do not contribute to the temperature adjustment which is mathematically described by the heat capacity coefficient. The phonon rates follow from a standard Boltzmann formula that can be found \textit{e.g.}~in Ref.~\cite{Pogna_acsnano_2022}. We remark that the coefficients discussed above depend on the electron distribution and phonon occupation, and must thus be calculated dynamically in time as the system evolves. Notwithstanding its simple appearance, equation~(\ref{eq:eom_temp}) is a strongly non-linear equation of motion.

The time derivative of the $\delta n_{\rm e}$ is given by the number of photons absorbed minus the number of phonons emitted by interband transitions, per unit time and area
\begin{equation}\label{eq:eom_dne}
\frac{d \delta n_{\rm e}(t)}{dt} = \frac{{\cal P}_{\rm CB}(t)}{\hbar \omega_{\rm CB}}+\frac{{\cal P}_{\rm FB}(t)}{\hbar \omega_{\rm FB}} - R_{\Gamma, {\rm inter}}(t) - R_{K, {\rm inter}}(t)~.
\end{equation}
Notice that photon absorption always results in an interband transition. We remark that the $\delta n_{\rm e}$ depends on interband phonon emission rate only, while the $T_{\rm e}$ depends an all phonon emissions: this is obviously because all phonon emissions reduce energy but only interband phonon emissions reduce the $\delta n_{\rm e}$.  

Finally, the rate equations for the phonon occupation are easily obtained by requiring consistency with Eqs.~(\ref{eq:eom_temp}) and~(\ref{eq:eom_dne}) in terms of energy and particle balance. Typical results of the integration of these rate equations are reported in the Supplementary Information S4.

\section{Discussion}
When a laser pulse is incident on a graphene flake, $T_{\rm e}$ increases over the pulse duration (see Supplementary Information S4) until it reaches a steady-state condition. In Ref.~\cite{soavi2018broadband} we safely used a steady-state condition in order to attribute the changes in THG signal to a single value of $T_{\rm e}$ for a fixed value of $E_{\rm F}$. This holds as long as one pulse measurement is performed on the graphene. In order to dedicate a single value to $T_{\rm e}$, either an instantaneous value or a value after the relaxation of the electrons (ps range) must be considered. However, considering the pulse durations used in our study of \qtyrange[range-units=single]{110}{150}{\femto\second}, limits us from both considerations. So this intermediate state in terms of pulse duration enables us to estimate a minimum and maximum for $T_{\rm e}$ for the experimental values in Fig.\;\ref{fig:modulation}. At $E_{\rm F} =$\;\SI{50}{\milli\electronvolt} and CB fluence of \qtylist[list-units=single]{11;200}{\micro\joule\per\square\centi\metre}, we estimate a $T_{\rm e}$ in the range $\sim$
\qtyrange[range-units=single]{1500}{1900}{\kelvin} and $\sim$ \qtyrange[range-units=single]{2300}{2500}{\kelvin}, respectively. At a higher value of doping ($E_{\rm F} =$\;\SI{300}{\milli\electronvolt}), we estimate a $T_{\rm e}$ in the range $\sim$ \;\qtyrange[range-units=single]{800}{1300}{\kelvin} for the CB fluence of \SI{11}{\micro\joule\per\square\centi\metre} and $T_{\rm e}$ $\sim$ \;\qtyrange[range-units=single]{2200}{2300}{\kelvin} for the CB fluence of \SI{200}{\micro\joule\per\square\centi\metre}. 

Furthermore, the origin of THG enhancement reported in Fig.\;\ref{fig:ambipolar} resulting from a reduction in $T_{\rm L}$ can be attributed to two coherent and incoherent physical processes. First, spectral broadening induced by FB leads to band broadening and alters carrier lifetimes, thereby affecting the THGE. Second, the well-established thermodynamics of carriers involving relaxation of carriers through optical phonons, which is temperature-dependent, contribute to the change in THGE.
In other words, the significant impact of $T_L$ on the TH modulation can be qualitatively understood based on two mechanisms, which include the dependence of electronic spectral broadening $\Gamma_e$ and kinetic relaxation rates $R(t)$ on lattice temperature $T_L$. The temperature dependence of $\Gamma_e$ predominantly originates from the scattering of electrons by acoustic phonons, while kinetic rates depend on temperature due to the electron-optical phonon interaction.

Finally, it is worth highlighting the interplay between the $T_e$ and photoexcited enhanced Pauli blocking. Steady-state theoretical considerations in Ref.~\cite{soavi2018broadband} predict that at low values of doping (when $E_{\rm F}<\hbar\omega/2$), increasing $T_{\rm e}$ will lead to the enhancement of the THG signal, a result that we were never able to observe experimentally in this work. However, these steady-state predictions rely on the assumption that $\delta n_{\rm e}$ remains constant once graphene is irradiated with a pulsed laser. In contrast, Fig.\;\ref{fig:efficiency} shows how the evolution of the {TH-MD} is accompanied by both the $T_{\rm e}$ and $\delta n_{\rm e}$ changes, both quantities that play a key role in the presence of both FB and CB, as discussed above. Thus, for instance, Fig.\;\ref{fig:efficiency}a shows the evolution of {TH-MD} when $E_{\rm F}/(\hbar \omega_{\rm FB})$ is 0.16. For lower values of doping (corresponding to $T_{\rm e}$ $\sim$ \qtyrange[range-units=single]{1500}{2500}{\kelvin} in our experiments) and ($\delta n_{\rm e}<$\;\num[print-unity-mantissa=false]{1e12}\;\unit[per-mode=reciprocal]{\per\centi\metre\squared}), {TH-MD} is always negative. This indicates that $\delta n_{\rm e}$ is not large enough to compete with the high $T_{\rm e}$, which is consistent with the experimental observations in Fig.\;\ref{fig:modulation}. On the other hand, when $E_{\rm F}/(\hbar \omega_{\rm FB})$ is 0.62, negative {TH-MD} occurs for $T_{\rm e} >$\;\SI{1300}{K} (Fig.\;\ref{fig:efficiency}b). Considering the $T_{\rm e}$ that we reach during the experiments (\qtyrange[range-units=single]{1500}{1900}{\kelvin}) at this regime of doping, {TH-MD} is still negative. This also confirms that $\delta n_{\rm e}$ in our experiments never reaches more than \num[print-unity-mantissa=false]{1e12}\;\unit[per-mode=reciprocal]{\per\centi\metre\squared}, where {TH-MD} would turn positive. It is worth mentioning that by comparing Fig.\;\ref{fig:efficiency}a and b, one can immediately notice that the change in {TH-MD} as a function of $T_{\rm e}$ is smaller when $E_{\rm F}/(\hbar \omega_{\rm FB})$ is 0.62. This behavior is consistent with the results in Ref.~\cite{soavi2019hot}. Therefore the THG in graphene is always accompanied by the two competing and interconnected effects of $T_{\rm e}$ (hot electrons) and $\delta n_{\rm e}$ (Pauli blocking).

\section{Conclusion}
In conclusion, we performed a detailed experimental and theoretical study of static thermal and ultrafast opto-electronic modulation of TH in a high-quality graphene FET encapsulated in thin hBN layers. 
As the main result of this study, we have established all-optical ultrafast control of graphene THG and achieved up to 85\% ultrafast opto-electronic modulation depth of the TH at $E_{\rm F}=$\;\SI{300}{\milli\electronvolt} and fluence of \SI{200}{\micro\joule\per\square\centi\metre}. 
Furthermore, this study addresses the static switching of THG \textit{via} tuning of the lattice temperature and electron doping. In particular, we measured that tuning of $T_{\rm L}$ from room temperature to \SI{33}{\kelvin} leads to a factor of $\sim$\;1.5 modulation of the $P_{TH}$ at $E_{\rm F}=$\;\SI{50}{\milli\electronvolt} and of $\sim$\;3 at $E_{\rm F}=$\;\SI{300}{\milli\electronvolt}. We suggest that his result originates from the spectral relaxation and thermodynamic kinetics of carriers. 
We discuss the $E_{\rm F}$ dependent temporal dynamics of all-optical TH modulation due to quenching of the phase-space scattering between optical phonons and electrons (\cite{pogna2021hot}). This provides a powerful tool to actively control both the TH modulation depth and the recombination dynamics in graphene opto-electronic nonlinear devices. 
Finally, we have addressed the experimental observations with a detailed theoretical framework that explains the ultrafast opto-electronic modulation of TH in graphene to be rooted in a mixed effect of Pauli blocking and carrier electronic temperature. Thus, this work provides a detailed description of the transient nonlinear optical and electronic response of graphene, which is crucial for the design of nanoscale and ultrafast optical modulators, detectors and frequency converters

\section{ACKNOWLEDGMENTS}
A.T.~acknowledges the ``National Centre for HPC, Big Data and Quantum Computing'', under the National Recovery and Resilience Plan (NRRP), Mission 4 Component 2 Investment 1.4 funded from the European Union - NextGenerationEU. 
H.R.~acknowledge the support from the Swedish Research Council (VR Starting Grant No. 2018-04252).
K.W. and T.T.~acknowledge support from the JSPS KAKENHI (Grant Numbers 21H05233 and 23H02052) and World Premier International Research Center Initiative (WPI), MEXT, Japan.
G.S.~acknowledges the German Research Foundation DFG (CRC 1375 NOA), project number 398816777 (subproject C4) and the International Research Training Group (IRTG) 2675 “Meta-Active”, project number 437527638 (subproject A4).

\end{document}


\section*{SUPPLEMENTARY INFORMATION}
Ultrafast Opto-Electronic and Thermal Tuning of Third-Harmonic Generation in a Graphene Field Effect Transistor


\section*{Author list}
Omid Ghaebi$^{1}$, Sebastian Klimmer$^{1,2}$, Nele Tornow$^{1}$, Niels Buijssen$^{1}$, Takashi Taniguchi$^{3}$, Kenji Watanabe$^{4}$, Andrea Tomadin$^{5}$, Habib Rostami$^{6}$, Giancarlo Soavi$^{1,7}$

\section*{Affiliations}
$^1$Institute of Solid State Physics, Friedrich Schiller University Jena, Jena, Germany

$^2$ARC Centre of Excellence for Transformative Meta-Optical Systems, Department of Electronic Materials Engineering, Research School of Physics, The Australian National University, Canberra, Australia

$^3$Research Center for Materials Nanoarchitectonics, National Institute for Materials Science,  1-1 Namiki, Tsukuba 305-0044, Japan

$^4$Research Center for Electronic and Optical Materials, National Institute for Materials Science, 1-1 Namiki, Tsukuba 305-0044, Japan

$^5$Dipartimento di Fisica, Università di Pisa, Largo Bruno Pontecorvo 3, 
56127 Pisa, Italy

$^6$Department of Physics, University of Bath, Claverton Down, Bath BA2 7AY, United Kingdom

$^7$Abbe Center of Photonics, Friedrich Schiller University Jena, Jena, Germany

\section{S1 Device fabrication}
\label{sec:sample fabrication}

The monolayer graphene flake was exfoliated from bulk synthetic graphite (HQ-graphene) using Scotch tape (Minitron). The hBN layers and graphite contacts were exfoliated with the same method on silicon wafers. The thickness of the hBN layers were determined by optical contrast following the approach described in Ref.~\cite{anzai2019broad}. A thin stamp comprising PC (polycarbonate) on a glass slide was prepared~\cite{purdie2018cleaning} an subsequently used to pick up the hBN layers, graphite contacts, and graphene using a commercial transfer stage (HQ-graphene). Subsequently, the layers were transferred to a silicon wafer (\SI{90}{\nano\metre} $\text{SiO}_{2}$) with pre-patterned gold contacts.

\section*{S2 Raman characterization and carrier mobility estimation}

\begin{figure*}
\centering
\includegraphics{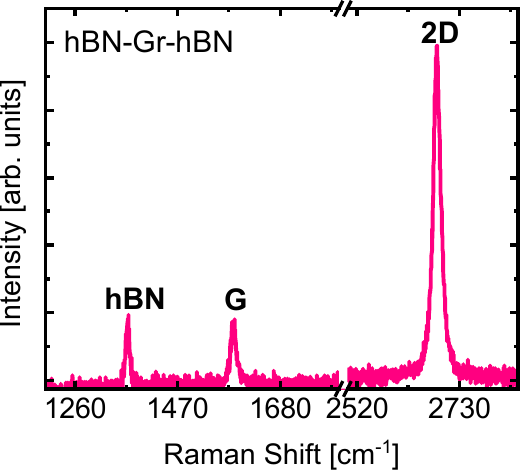}
\caption{\label{fig:raman}
\textbf{Raman characterization.} Raman spectrum of the device showing the 2D peak of graphene at \SI[per-mode=reciprocal]{2683}{\per\centi\metre}, the G peak of graphene at \SI[per-mode=reciprocal]{1596}{\per\centi\metre} and the $E_{\rm 2g}$  mode of hBN located at \SI[per-mode=reciprocal]{1368}{\per\centi\metre}.}
\end{figure*}
 
After fabrication, we characterized the device with Raman spectroscopy. The Raman spectrum of the sample after hBN encapsulation is depicted in Fig.\;\ref{fig:raman}a. The 2D peak at \SI[per-mode=reciprocal]{2683}{\per\centi\metre} is a single Lorentzian, confirming the monolayer nature of our sample~\cite{ferrari2006raman}. The low FWHM (Full Width at Half Maximum) of the G and 2D peaks (FWHM(G)\;=\;\SI[per-mode=reciprocal]{13.5}{\per\centi\metre}, FWHM(2D)\;=\;\SI[per-mode=reciprocal]{17.7}{\per\centi\metre} indicates a negligible strain of the graphene flake after transfer~\cite{neumann2015raman}. The peak at \SI[per-mode=reciprocal]{1368}{\per\centi\metre} belongs to the hBN bottom and top flakes (in-plane atom vibrations)\cite{schue2016characterization,geick1966normal}. The absence of the D peak of graphene, typically at \SI[per-mode=reciprocal]{1350}{\per\centi\metre}, is a further indication of the high quality of our sample~\cite{tuinstra1970raman,ferrari2000interpretation,ferrari2013raman}. 

Further, we estimated the mobility of the device from the black $I_{\rm SD}-V_{\rm G}$ curve in Fig.\;1b of the main text. The carrier mobility is proportional to the first derivative of $I_{\rm SD}$ with respect to $V_{\rm G}$ \text{via} equation~(\ref{eq:mobility})~\cite{purdie2018cleaning,de2019high}:
\begin{equation}\label{eq:mobility}
     \mu =   \frac{1}{c_{\rm eff}} \   \frac{d\sigma}{dV_{\rm G}} =  \frac{L}{\rm w\epsilon} \  \ \frac{d_{c}}{v_{\rm SD}} \frac{dI_{\rm SD}}{dV_{\rm G}}
\end{equation}
where $L$ and $w$ define the size of the channel (graphene flake), $\epsilon$ is the relative permittivity, and $c_{\rm eff}$ is the effective capacitance of the dielectric spacer (hBN/$\text{SiO}_{\rm 2}$). The thickness of hBN is $\sim$\;\SI{12}{\nano\metre} (identified by color based on ~Ref.\cite{anzai2019broad}) and the oxide thickness is \SI{90}{\nano\metre}. Therefore, $c_{\rm eff}$ can be calculated considering hBN and oxide layers in series~\cite{purdie2018cleaning}. From this, we obtained a mobility of $\sim$\;\SI{25000}{\centi\metre\squared\per{\volt\second}}.

In a graphene FET, $V_{\rm G}$ creates an electrostatic potential between graphene and the gate electrode which can tune the $E_{\rm F}$ by changing the electron density $n_{\rm e}$. As a result, $V_{\rm G}$ is given by :
\begin{equation}
     V_{\rm G}= \frac{E_{\rm F}}{e}+\varphi 
\end{equation}
The first and second terms are determined by the quantum and geometrical (effective) capacitance ($c_{\rm eff}$), respectively. For a back-gated sample, the geometrical capacitance dominates over the quantum capacitance: $V_{\rm G}\sim\varphi=\frac{n_{\rm e}e}{c_{\rm eff}}$ ~\cite{das2008monitoring}. In graphene, the $E_{\rm F}$ is proportional to the square root of $n_e$ \textit{via} the relation $E_{\rm F}= \hbar v_{\rm F}\sqrt n_{\rm e}$ with $v_{\rm F}\sim$\;\SI[print-unity-mantissa=false]{1e6}{\metre\per\second}~\cite{novoselov2005two}. Considering $c_{\rm eff}=\frac{\epsilon_{\rm r}\epsilon_{0}}{d_{c}}$, $E_{\rm F}$ can be written as: 

\begin{equation}
     E_{\rm F}= \hbar v_{\rm F}\sqrt\frac{\pi\epsilon_{0}\epsilon_{\rm r} (V_{\rm G}-V_{\rm CNP})}{ed_{c}}
\end{equation}

where $\epsilon_{\rm r}$ and $d_{c}$ are the relative permittivity and thickness of the  capacitor and $V_{\rm CNP}$ is the gate voltage at minimum conductance (Charge Neutrality Point, CNP). 

\section{S3 THG experimental setup}
\label{sec:experimental details}

The CB and FB pulses were obtained from a Yb-based femtosecond oscillator (FLINT12, Light Conversion) and OPO (Levante fs IR, APE), respectively. The main laser source operates at \SI{76}{\mega\hertz} and \SI{12}{\watt} of average power. A portion of this (\SI{4.5}{\watt}) is used to pump the OPO, which provides tunable output in the range \qtyrange[range-units=single]{1320}{2000}{\nano\metre} for the signal and \qtyrange[range-units=single]{2150}{4800}{\nano\metre} for the idler. The pulse duration of FB and CB are \qtylist[list-units=single]{150;110}{\femto\second}, respectively. The relative delay between CB and FB is controlled by a motorized delay line (M-404.2PD, PI). The two pulses are subsequently combined on a beam splitter (BS), after which they propagate collinearly into a home-built microscope and they are finally simultaneously focused on the sample with a spot size of $\sim$\;\SI{6.7}{\micro\metre} (FB) and $\sim$\;\SI{2.2}{\micro\metre} (CB), measured from the razor blade technique~\cite{kimura1987method}. In all the experiments, the sample was mounted inside an optical cryostat (ST-500, Janis) coupled to a silicon temperature controller (Lakeshore) integrated with built-in stages (Attocube, ANPX101/LT and ANPZ102/RES/LT). The backward emitted THG is spectrally filtered and detected on an amplified InGaAs photoreceiver (model 2153, Newport). The THG signal experiences a total loss of 92.8\% in propagation through different optical components and considering the quantum efficiency of the detector. The peak fluence of the FB is kept at $\sim$\;\SI{130}{\micro\joule\per\square\centi\metre} for static and at $\sim$\;\SI{110}{\micro\joule\per\square\centi\metre} for all-optical modulation THG experiments, respectively, while the CB peak fluence is tuned in the range \qtyrange[range-units=single]{11}{200}{\micro\joule\per\square\centi\metre}.

\begin{figure}
\includegraphics[width=\linewidth]{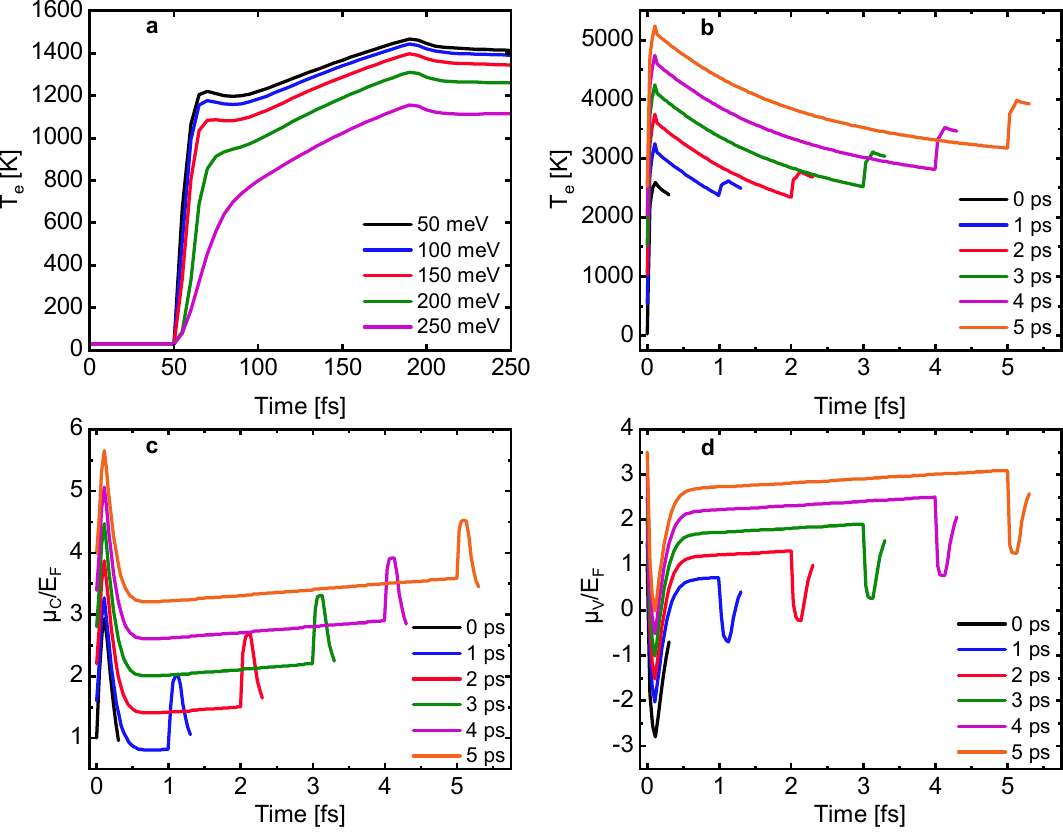}
\caption{\label{fig:thermodynamics}
\textbf{Time-evolution of $T_{\rm e}$, $\mu_{\rm C}$, and $\mu_{\rm V}$}. Time-evolution of $T_{\rm e}$ with \textbf{a} a single pulse and several values of $E_{\rm F}$ and with \textbf{b} two pulses as the delay is increased.
\textbf{c,d} Time-evolution of $\mu_{\rm C} / E_{\rm F}$ and $\mu_{\rm V} / E_{\rm F}$ \textbf{d} for the same conditions as \textbf{b}. The plot legends indicate the time delay between FB and CB pulses. }
\end{figure}

\section{S4 Time-evolution of the electron distribution}
\label{sec:examples_evolution}

Fig.\;\ref{fig:thermodynamics} shows the time-evolution of electron thermodynamic variables for a set of typical parameters.
With reference to the model described in the main text, we use a typical integration timestep t $\sim$\;\SI{5}{\femto\second} and refractive indices of the ${\rm Si}{\rm O}_{2}$ substrate $n_{\rm t},\,n_{\rm b}\;=\;1.4$, ignoring the thin hBN layers because excitations do not overlap with the frequency ranges where the material features hyperbolic dispersion.
The phonon rates $R_{\Gamma,{\rm K}}$, given \textit{e.g.}~in Ref.~\cite{Pogna_acsnano_2022} are proportional to the electron-phonon coupling parameter $\partial t/ \partial b$, see also Refs.~\cite{rana2009carrier,wang2010ultrafast}.
We use the value $\partial t/ \partial b =$\;\SI{200}{\electronvolt\per{\nano\metre}}, substantially larger than found in the literature, to address at a phenomenological level the enhanced electron recombination that has been reported in Refs.~\cite{pogna2021hot, Pogna_acsnano_2022}.

In (a) the electron temperature is shown for several values of $E_F$, in the absence of a CB, under a FB active from $t =$\;\SI{50}{\femto\second}, of duration \SI{150}{\femto\second} and fluence \SI{130}{\micro\joule\per\square\centi\metre}. Temperatures of the order of \SI[print-unity-mantissa=false]{1e3}{\kelvin} are achieved during the FB.
As expected, a higher $E_F$ leads to lower $T_e$, because a higher carrier density entails a larger heat capacity.

Panels (b)-(d) correspond to a dynamics including a CB active from $t = 0$, of duration \SI{110}{\femto\second} and fluence \SI{200}{\micro\joule\per\square\centi\metre}, and a FB of duration \SI{150}{\femto\second} and fluence \SI{110}{\micro\joule\per\square\centi\metre}, active after an initial delay shown in the legend.
The curves are displaced along the vertical axis for visibility.

In particular, (b) shows that a FB with same fluence leads to markedly different final temperature, based on the initial condition of the electron system.
One also has to keep in mind that the absorption coefficient in graphene does depend on temperature and the chemical potentials, and thus changes with the delay.
It is also important to notice that $T_e$ is not constant during the FB, even if a smaller variation is experienced compared to case (a), where the electron system is in equilibrium at the lattice temperature of \SI{30}{\kelvin} before the FB. 
Finally, the role of the photoexcited density is clearly visible in (d), where a change of sign of the chemical potential in the valence band takes place during the CB and FB, corresponding to a large quantity of holes being left behind by the electrons promoted to the conduction band.

We reiterate that all these results are obtained assuming the quasi-equilibrium form (equation~(9) of the main text) for the electron distribution, and can only be understood in the sense of a coarse-grained representation of the time-evolution, on a time-step larger than the thermalization time-scale of $\sim$\;\SI{20}{\femto\second}.

\section{S5 Finite conductivity and the electron scattering rate}
\label{sec:scatteringrate}

The electron scattering rate $\Gamma_{\rm e}$ that enters equation~(6) of the main text can be expressed as the sum of scattering rates from different sources: $\Gamma_e = (\Gamma_e)_{\rm ac} + (\Gamma_e)_{\rm imp}$.
Refs.~\cite{Ando1998,Ando2006,DasSarma_rmp_2011} provide formulas for the electron scattering rate due to scattering from long-range charged impurities and short-range disorder, given by $(\Gamma_e)_{\rm imp} = (\Gamma_e)_{\rm long} + (\Gamma_e)_{\rm short}$:
\begin{align}
(\Gamma_e)_{\rm long}  & \approx  \frac{n_{\rm i} (\pi r_{\rm s})^2 }{2} \left \{\frac{(\hbar v_{\rm F})^2}{|E_{\rm F}|}\right\},
\\
(\Gamma_e)_{\rm short} &\approx \frac{n_{\rm d} V^2_0}{8} \left\{\frac{|E_{\rm F}|}{(\hbar v_{\rm F})^2}\right\},
\end{align}
where $n_{\rm i}$ is charged impurity center density, $r_{\rm s}=e^2/(\hbar v_{\rm F} \kappa)$ with $\kappa$ being the dielectric constant, $n_{\rm d}$ is the short-range impurity density and $V_0$ is a constant short-range  potential strength.
For low $T_{\rm L}<$\;\SI{200}{\kelvin}, the resistivity of graphene is primarily influenced by scattering with acoustic phonons, contributing to the electron scattering rate at the Fermi surface as follows~\cite{Hwang_prb_2008}
\begin{align}
(\Gamma_e)_{\rm ac} \approx \frac{D^2}{8\rho_m v_{\rm s}^2} \left\{\frac{|E_{\rm F}|}{(\hbar v_{\rm F})^2}\right\} (k_B T_L), 
\end{align}
where $D=$\;\SI{19}{\electronvolt} is the deformation potential, $v_{\rm s}\sim$\;\SI{2e6}{\centi\metre\per\second} is the sound velocity, and $\rho_{\rm m}=$\;\SI{7.6e-8}{\gram\per\centi\metre\squared} is the mass density.

Taking into account contributions from acoustic phonons, long-range charged impurities, and short-range disorder, we can use the following empirical ansatz for the electron scattering rate:
\begin{align}
\Gamma_{\rm e} = A |E_{\rm F}| + \frac{B}{|E_{\rm F}|} + C |E_{\rm F}| (k_B T_L),
\end{align}
where $A$, $B$, and $C$ are empirical parameters and in quasi-equilibrium condition we can replace $E_F$ with the average of chemical potential in the conduction and valence bands: $E_{\rm F} \to (\mu_{\rm C}+\mu_{\rm V})/2$.  
By relating the electron scattering rate to the mobility and fitting the mobility \textit{vs.} $E_F$ to experimental data we find $B =$\;\SI{0.0013}{\electronvolt\squared}. At present, we neglect the coefficients $A$ and $C$ for simplicity.